\documentclass[showpacs,preprintnumbers,amsmath,amssymb]{revtex4}

\usepackage{graphicx}
\usepackage{dcolumn}
\usepackage{bm}
\begin{document}
\title{Neutral Current induced $\pi^0$ production and neutrino magnetic moment}
\author{M. Sajjad Athar, S. Chauhan and S. K. Singh}
\affiliation{Department of Physics, Aligarh Muslim University, Aligarh-202 002, India}
\date{\today}
\begin{abstract}
We have studied the total cross section, $Q^2$, momentum and angular distributions for pions in the $\nu$($\bar \nu$) induced $\pi^0$ production from nucleons. The calculations have been done for the weak production induced by the neutral current in the standard model and the electromagnetic production induced by neutrino magnetic moment. It has been found that with the present experimental limits on the muon neutrino magnetic moment $\mu_{\nu_\mu}$, the electromagnetic contribution to the cross section for the $\pi^0$ production is small. The neutrino induced neutral current production of $\pi^{0}$, while giving an alternative method to study the magnetic moment of neutrino $\mu_{\nu_\mu}$, does not provide any improvement over the present experimental limit on $\mu_{\nu_\mu}$ from the observation of this process in future experiments at T2K and NO$\nu$A.
\end{abstract}
\maketitle
\section{Introduction}
The neutral current $\pi^0$ production in neutrino interactions plays an important role in the background studies of $\nu_\mu \rightarrow \nu_e$ oscillations in the appearance mode as well as in discriminating between $\nu_\mu \rightarrow \nu_\tau$ and $\nu_\mu \rightarrow \nu_s$ modes~\cite{Aguiler1, Kajita, Mauger, Raaf}. This process can also help to distinguish between production of $\nu_\tau$ and $\bar \nu_\tau$ in some oscillation scenarios at  neutrino energies much below the $\tau$ production threshold but above the pion threshold~\cite{Nieves}. The recent results on neutral current induced pion production in neutrino oscillation experiments at K2K~\cite{aip, nakayama} and MiniBooNE~\cite{Aguiler, nguyen, link} have generated great interest in studying these processes. In this context, the proposed experiments by T2K~\cite{Yury, nakadaira} and NO$\nu$A~\cite{norman} collaborations plan to study this process with better statistics. 

The neutral pions can also be produced by electromagnetic interactions if $\nu(\bar \nu)$ have diagonal and/or transition magnetic moments. This process would in principle contribute additional events to the neutral current reaction and would modify the energy and angular distributions of the neutral pions, which may be observed in future experiments. It is thus possible, in principle, to get information about the magnetic moment of neutrinos(antineutrinos) from studying neutral current induced $\pi^{0}$ production from nucleons and nuclei. While the minimal extensions of Standard model predict very tiny diagonal magnetic moments~\cite{Fujikawa}, there are models of electroweak interactions which predict enhanced transition magnetic moment~\cite{Mohapatra}. The present limits for the magnetic moment of neutrinos come from neutrino - electron scattering for $\nu_{\mu}$ and from $e^{+}e^{-} \rightarrow \nu \bar \nu \gamma$ for $\nu_{\tau}$. These limits for $\nu_\mu$ and $\nu_\tau$ magnetic moments are  $\mu_{\nu_{\mu}} < 6.8 \times 10^{-10} \mu_{B}$ and $\mu_{\nu_{\tau}} < 3.9 \times 10^{-7} \mu_{B}$~\cite{review}.

 The data from neutrino oscillation experiments from Sudbury Neutrino Observatory and Super-Kamiokande have also been analysed to obtain improved limits on the neutrino magnetic moments for $\nu_{\mu}$ and $\nu_{\tau}$~\cite{grifols, grimus, joshipura, beacon, pulido}. A recent analysis of the Borexino experiment claims to improve these limits on the magnetic moments of $\nu_{\mu}$ and $\nu_{\tau}$ by 3 orders of magnitude\cite{montanino}.  

We would like to consider the possibility of obtaining new bounds on the neutrino magnetic moment using high statistics data on neutral current induced $\pi^{0}$ production from nucleons and nuclei in future experiments. Such a possibility was earlier discussed by Kang et al.~\cite{kang} using first result on $\pi^{0}$ production from Superkamiokande experiment on atmospheric neutrinos\cite{Fukuda}.

We study in this paper, the $\pi^{0}$ production induced by weak neutral current and magnetic moment interaction of neutrinos and antineutrinos in the energy region of few GeV, relevant for K2K, MiniBooNE, T2K and NO$\nu$A experiments.

In Sec.II, we give the formalism and present our results for the total cross section $\sigma$, $Q^2$ distribution ($\frac{d\sigma}{dQ^2}$), momentum distribution ($\frac{d\sigma}{dp_{\pi}}$) and angular distribution($\frac{d\sigma}{dcos\theta_{\pi q}}$) for $\pi^{0}$ in Sec.III,  where we also discuss the possibility of obtaining improved limits on $\nu(\bar \nu)$ magnetic moments. 
\section{Formalism}
 In the energy region of one GeV relevant for atmospheric neutrinos and present accelerator neutrino experiments the dominant process of pion production is through the excitation of $\Delta$ resonance and its subsequent decay to pions, i.e.  
\begin{equation} \label{reaction}
\nu N\rightarrow \nu \Delta \rightarrow \nu N \pi^{0}.
\end{equation}
\begin{figure}
\includegraphics{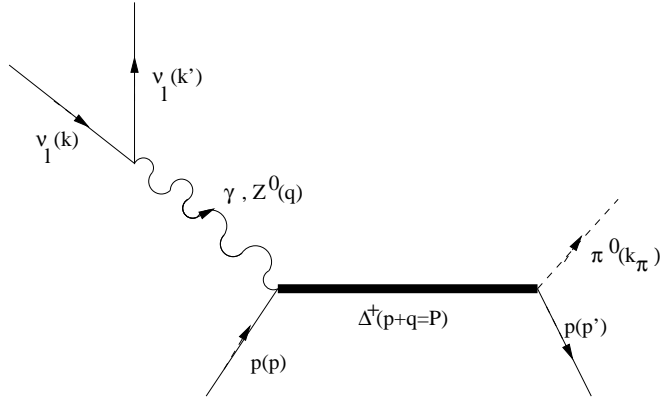}
\caption{Feynman diagram for the process $\nu + p(p)\rightarrow \nu(k^\prime)+p(p^{\prime})+\pi^{0}(k_{\pi})$.}
\end{figure}
The differential scattering cross section for the reaction $\nu(k)+p(p)\rightarrow \nu(k^\prime)+p(p^{\prime})+\pi^{0}(k_{\pi})$ shown in Fig.1. is given by 
\begin{equation} \label{cross_section}
d \sigma = \frac{(2\pi)^4 \delta^{4}(p_{i} - p_{f})}{4 {p \cdot k}} \prod_{j=1}^{3}\frac{d^{3}p_{j}}{(2\pi)^{3} 2 E_{j}}|\mathcal M_{fi}|^2,
\end{equation}
where $p_{i}(=k+p)$ and $p_{f}(=\sum_{j=1}^{3} p_{j})$ are the four momenta of the initial and final states, respectively. The transition matrix element $\mathcal M_{fi}$ is written using 
\begin{equation}
{\it L}^{W}=\frac{G_{F}}{\sqrt{2}} l_{\alpha}j^{\alpha}
\end{equation}
 for the weak $Z N\Delta$ interaction and 
\begin{equation}
 {\it L}_{\pi N \Delta}= \frac{f_{\pi N \Delta}}{m_{\pi}}{\bar\Psi}_\mu {\vec T}^\dagger (\partial^\mu{\vec \phi})\Psi + h. c.
\end{equation}
 for the strong  $\pi N\Delta$ interaction. $\Psi_\mu$ is a Rarita-Schwinger field for spin-$\frac{3}{2}$ particle, ${\vec T}^\dagger$ is the isospin transition operator, $\vec \phi$ is the pion field.

 The matrix element of the leptonic current $l_{\alpha}$  and the hadronic current $j^{\alpha}$ are defined as 
\begin{equation}
<k|l_{\alpha}|k^{\prime}>=\bar u({\bf k}^{\prime}) \gamma_{\alpha}(1-\gamma^{5})u({\bf k}),
\end{equation}
and
\begin{equation}
<\Delta(P)|j^\alpha|p>= \bar \Psi_{\beta}({\bf P}){\bf \mathcal O}^{\beta \alpha} u({\bf p})
\end{equation}
$u({\bf p})$ is the Dirac spinor for the proton. 

$\mathcal O^{\beta \alpha}=(1-2 sin^{2}\theta_{W}){\mathcal O}_V^{\beta \alpha}+{\mathcal O}_A^{\beta \alpha}$ for the neutral current process with ${\mathcal O}_V^{\beta \alpha}$ and ${\mathcal O}_A^{\beta \alpha}$ given by
\begin{eqnarray}\label{vec_tra_current}
{\mathcal O}_V^{\beta \alpha}&=\left(\frac{C_{3}^V(q^2)}{M}(g^{\alpha \beta} \not q-q^{\beta}\gamma^{\alpha})+\frac{C_{4}^V(q^2)}{M^2}(g^{\alpha \beta} q \cdot P-q^{\beta}P^{\alpha})+\frac{C_{5}^V(q^2)}{M^2}(g^{\alpha \beta}q \cdot p-q^{\beta}p^{\alpha})\right)\gamma_{5}
\end{eqnarray}
and
\begin{eqnarray}\label{ax_tra_current}
{\mathcal O}_A^{\beta \alpha}&=&\left(\frac{C_{4}^{A}(q^2)}{M^2}(g^{\alpha \beta}{\not q} -q^{\beta}\gamma^{\alpha})+C_{5}^{A}(q^2)g^{\alpha \beta}
+\frac{C_{6}^{A}(q^2)}{M^2}q^{\beta}q^{\alpha}\right)
\end{eqnarray}
where ~$C_{i}^{V}(q^2)$ and  $C_{i}^{A}(q^2)$ are the vector and axial vector transition form factors, and $\theta_{W}$ is the Weinberg angle ($sin^{2}\theta_{W}$= 0.23122). q(=$k-k^\prime$) is the four momentum transfer.  M is the mass of the nucleon.

Using Eqs. (3)-(6) the matrix element for the process $\nu + p\rightarrow \nu+p+\pi^{0}$ in the $\Delta$ dominance model is written as 
\begin{equation}\label{matrix_element}
\mathcal M_{fi}=\sqrt{\frac{2}{3}} \frac{G_F}{\sqrt{2}}\frac{f_{\pi N \Delta}}{m_{\pi}} \bar u({\bf p}^{\prime}) k^{\sigma}_{\pi} {\mathcal P}_{\sigma \lambda} \mathcal O^{\lambda \alpha} l_{\alpha} u({\bf p})
\end{equation}
where $\sqrt{\frac{2}{3}}$ has appeared because of the isospin factor coming at the vertex $\Delta^{+}\rightarrow p \pi^{0}$. $G_{F}(=1.16637\times 10^{-5} GeV^{-2})$ is the Fermi coupling constant.

${\mathcal P}^{\sigma \lambda}$ is the $\Delta$ propagator in momentum space given by 
\begin{equation}	\label{width}
{\mathcal P}^{\sigma \lambda}=\frac{{\it P}^{\sigma \lambda}}{P^2-M_\Delta^2+iM_\Delta\Gamma},
\end{equation}
where ${\it P}^{\sigma \lambda}$ is the spin-3/2 projection operator given by
\begin{eqnarray}\label{propagator}
{\it P}^{\sigma \lambda} = \sum_{spins} \psi^{\sigma} \bar \psi^{\lambda} = (\not P+M_{\Delta})
\left(g^{\sigma \lambda}-\frac{2}{3} \frac{P^{\sigma}P^{\lambda}}{M_{\Delta}^2}+\frac{1}{3}\frac{P^{\sigma} \gamma^{\lambda}-P^{\sigma} \gamma^{\lambda}}{M_{\Delta}}-\frac{1}{3}\gamma^{\sigma}\gamma^{\lambda}\right),
\end{eqnarray}
and the delta decay width $\Gamma$ is taken to be an energy dependent P-wave decay width taken as
\begin{equation}
\Gamma(W)=\frac{1}{6 \pi}\left(\frac{f_{\pi N \Delta}}{m_{\pi}}\right)^2 \frac{M}{W}|q_{cm}|^3.
\end{equation}
$|q_{cm}|$ is the pion momentum in the rest frame of the resonance and is given by
\[|q_{cm}|=\frac{\sqrt{(W^2-m_{\pi}^2-M^2)^2 -4 m_{\pi}^2M^2}}{2W},\]
with W as the center of mass energy.

If the reaction shown in Eq.(1) is induced by neutrino magnetic moment then the matrix element given by Eq.(9) would modify to 
\begin{equation}\label{matrix_element}
\mathcal M_{fi}=\sqrt{\frac{2}{3}} \frac{f_{\pi N \Delta}}{m_{\pi}} \bar u({\bf p}^{\prime}) k^{\sigma}_{\pi} {\mathcal P}_{\sigma \lambda} \mathcal O^{\lambda \alpha}_{V} l_{\alpha}^{em} u({\bf p}),
\end{equation}
where
\begin{equation}
l_{\alpha}^{em}=\mu_\nu^{eff} \bar u({\bf k}^{\prime}){\frac{\sigma^{\alpha \beta}}{q^2}} q_{\beta} u({\bf k}),
\end{equation}
with $\mu_{\nu}^{eff}$ as the effective magnetic moment of the neutrino, which is given in terms of the magnetic moments of the mass eigen states and oscillation probabilities that depend upon the specific oscillation  models used for analysing the neutrino oscillation experiments \cite{grifols, grimus, joshipura, beacon, pulido}. 
\begin{figure}
\includegraphics{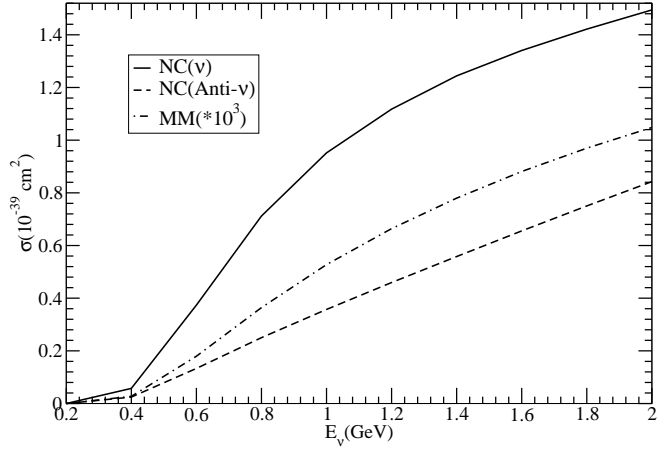}
\caption{Total scattering cross section for the reaction $\nu+p \rightarrow \nu + p+ \pi^0$ induced by weak neutral current and the magnetic moment induced processes.}
\end{figure}
\begin{figure}
\includegraphics{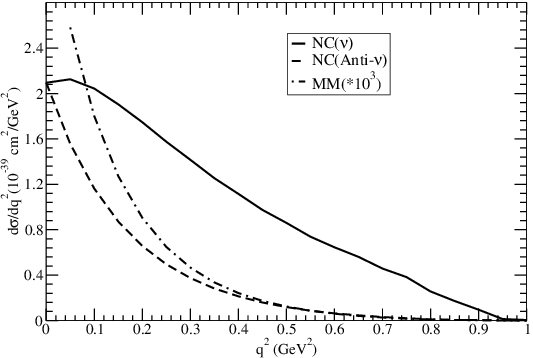}
\caption{$Q^2$ distribution for the weak neutral current and the magnetic moment induced processes at $E_{\nu}=1 GeV$.}
\end{figure}
\begin{figure}
\includegraphics{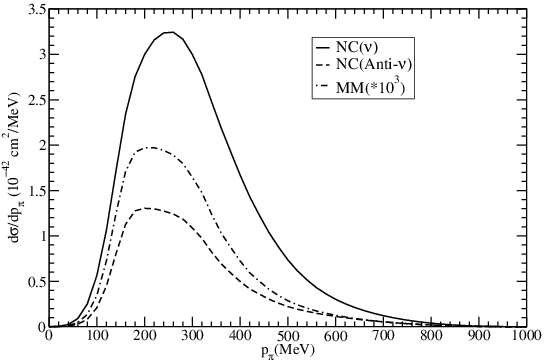}
\caption{Pion momentum distribution for the weak neutral current and the magnetic moment induced processes at $E_{\nu}=1 GeV$.}
\end{figure}
\begin{figure}
\includegraphics{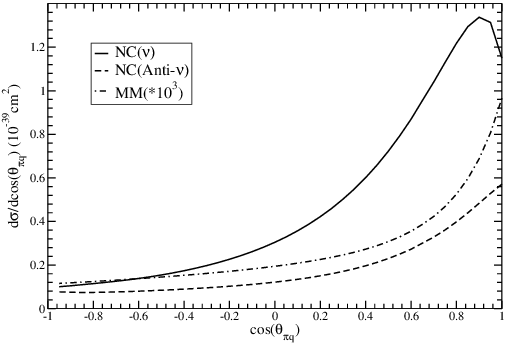}
\caption{Pion angular distribution for the weak neutral current and the magnetic moment induced processes at $E_{\nu}=1 GeV$.}
\end{figure} 
We have taken the following form of the $N-\Delta$ transition form factors \cite{lalakulich} 
\begin{eqnarray}\label{vec_ff}
C_i^V(Q^2)&=&C_i^V(0)~(1+\frac{Q^2}{M_V^2})^{-2}~{D_{i}}\nonumber\\
D_{i}&=&(1+\frac{Q^2}{4M_V^2})^{-1};~~i=3,4 \nonumber\\
D_{i}&=&(1+\frac{Q^2}{0.776M_V^2})^{-1};~~i=5
\end{eqnarray}
\[C_{3}^{V}(0)=2.13, C_{4}^{V}(0)=-1.51,\]\[ C_{5}^{V}(0)=0.48\]
where $M_V(=0.84 GeV)$ is the vector dipole mass.

The axial vector form factors are parametrised as
\begin{eqnarray}\label{axi_ff}
C_i^A(Q^2)&=&C_i^A(0)~(1+\frac{Q^2}{M_A^2})^{-2}~{D_{i}}\nonumber\\
D_{i}&=&(1+\frac{Q^2}{3M_A^2})^{-1};
\end{eqnarray}
\[C_{4}^{A}(Q^2)=-\frac{C_{5}^{A}(Q^2)}{4},C_{5}^{A}(0)=1.2,\]\[ C_{6}^{A}(Q^2)=C_{5}^{A}(Q^2)\frac{M^2}{m_{\pi}^{2}+Q^2}\]
where $M_{A}(=1.05 GeV)$ is axial vector dipole mass.

The differential scattering cross section $\frac{d^5 \sigma}{dQ^{2} d\Omega_{\pi}dp_{l}}$ is calculated using Eq.(2), and is written as 
\begin{eqnarray}
\frac{d^5 \sigma}{dQ^{2} d\Omega_{\pi} dp_{l}} = \frac{1}{(4\pi)^5} \frac{\pi}{E_{\nu} E_{l}}\frac{|\vec  k^{\prime}||\vec k_{\pi}|}{M E_{\nu}}
\frac{1}{E_{p}^{\prime}+E_{\pi}\left(1-\frac{|\vec q|}{|\vec k_{\pi}|}cos(\theta_{\pi q})\right)} \bar \sum \sum|\mathcal M_{fi}|^2
\end{eqnarray}
where $|\vec k_{\pi}|$ is the pion momentum.
Similarly we get an expression for the pion distribution using Eq. (2).
\section{Results and Discussions}
The numerical results for the total cross section $\sigma$, the differential cross sections $\frac{d\sigma}{dQ^2}$, $\frac{d\sigma}{dcos\theta_{\pi q}}$ and $\frac{d\sigma}{dp_{\pi}}$ for the neutral current production of $\pi^0$ induced by neutrinos(antineutrinos) are presented in Figs.2-5 along with the contributions of the electromagnetic production induced by the neutrino(antineutrino) magnetic moment. For the neutral current production, the numerical values of the vector and axial vector form factors given in Eqs.(\ref{vec_ff}) \& (\ref{axi_ff}) have been used while for the electromagnetic production the numerical values of the vector form factors given in Eq.(\ref{vec_ff})   along with the neutrino magnetic moment $\mu_{\nu}^{eff} = 6.8 \times 10 ^{-10} \mu_{B}$ is used.  A momentum dependent strong form factor with $f_{\pi N \Delta}(m_{\pi}^2)=2.12$ ~\cite{Penner} has been used in numerical calculations.

We show in Fig 2., the total cross section $\sigma$, for the neutral current induced $\nu(\bar \nu)$ production of $\pi^{0}$. The present results are in agreement with the results of Leitner et al.\cite{ruso} and also with the results of Hernandez et al.\cite{nieves} if their values for the parameter $C_{5}^{A}(0)$ and $M_{A}$ are used but are in disagreement with results of Kang et al.\cite{kang} who find a smaller value for $\sigma$. We also show in this figure, the total cross section $\sigma$ for electromagnetic production of $\pi^{0}$ induced by neutrino magnetic moment which is in agreement with the results of Kang et al.\cite{kang} if neutrino magnetic moment $\mu_{\nu}^{eff} = 6 \times 10 ^{-9} \mu_{B}$ as used by them is taken. We see in Fig. 2. that with the present limits on the magnetic moment of $\nu_{\mu}$, the electromagnetic production of $\pi^0$ is $10^{-3}$ times smaller than the neutral current induced $\pi^0$ production. It is, therefore, not feasible to improve the present limit on neutrino magnetic moment from $\pi^0$ production cross section measurements as earlier expected from the work of Kang et al.~\cite{kang}. 

In Figs. 3-5, we also show the differential cross sections $\frac{d\sigma}{dQ^2}$, $\frac{d\sigma}{dcos\theta_{\pi q}}$ and $\frac{d\sigma}{dp_{\pi}}$ for the neutral current induced $\pi^0$ production by $\nu$ and $\bar\nu$ as well as the $\pi^0$ production induced by magnetic moment of $\nu$($\bar\nu$). The present experiments at MiniBooNE \cite{Aguiler} see neutral current induced $\pi^0$ events of the order of $2.8 \times 10^4$ which can be further increased by an order of magnitude at T2K and NO$\nu$A. These pions are produced on nuclear targets like $^{12}C$. In the case of nuclear targets, there are incoherent as well as coherent production of $\pi^0$ which have different angular and momentum distributions. An analysis of these experiments in order to study the neutrino magnetic moment would require an understanding of nuclear effects in the incoherent and coherent production of $\pi^0$ induced by the neutral currents as well as by the neutrino magnetic moment on nuclear targets in the energy region of 1 GeV.

We would like to conclude that it is possible in principle to study the neutrino magnetic moment from the observations of neutral current induced $\pi^0$ production from nuclear targets in the near detector in future neutrino oscillation experiments by T2K \& NO$\nu$A collaborations. However, with the present limits on $\mu_{\nu_\mu}$ the magnetic moment induced $\pi^0$ production cross sections are quite smaller then the weak neutral current induced cross sections. It is, therefore, not a feasible method to constrain the neutrino magnetic moment beyond the present experimental limits.
\section{Acknowledgments} One of the authors (S.C.) is thankful to the Jawaharlal Nehru Memorial Fund for the Doctoral Fellowship.


\begin{references}
\bibitem{Aguiler1}  A. A. Aguilar-Arevalo et al., Phys, Rev. Lett. {\bf 98} 231801 (2007).
\bibitem{Kajita} T. Kajita, Nucl. Phys. B Proc. Suppl. {\bf 77} 123 (1999).
\bibitem{Mauger} C. M. Mauger, Ph.D. Thesis, SUNY, Stony Brook [UMI-30-88072, (2002)(unpublished)].
\bibitem{Raaf} J. L. Raaf, Ph.D. Thesis, Cincinnati U. [FERMILAB-THESIS-2005-20, (2005)(unpublished)].
\bibitem{Nieves} E. Hernandez, J. Nieves, and M. Valverde, Phys. Lett. B {\bf 647}, 452 (2007). 
\bibitem{aip} C. Mariani, A. I. P. Conf. Proc. {\bf 967}, 174 (2007).
\bibitem{nakayama} S. Nakayama et al., Phys. Lett.  B {\bf 619}, 255 (2005).
\bibitem{Aguiler} A. A. Aguilar-Arevalo et al., Phys. Lett.  B {\bf 664}, 41 (2008).
\bibitem{nguyen} V. T. Nguyen, A. I. P. Conf. Proc. {\bf 981}, 250 (2008).
\bibitem{link} J. M. Link,  A. I. P. Conf. Proc. {\bf 967}, 151 (2007).
\bibitem{nakadaira} T. Nakadaira, A. I. P. Conf. Proc. {\bf 981}, 222 (2008).
\bibitem{Yury} Y. Kudenko, physics.ins-det/0805.0411, (2008).
\bibitem{norman} A. Norman,  A. I. P. Conf. Proc. {\bf 981}, 225 (2008).
\bibitem{Fujikawa} K. Fujikawa, and R. Shrock, Phys. Rev. Lett. {\bf 45}, 963 (1980).
\bibitem{Mohapatra} R. N. Mohapatra and P. B. Pal, ``Massive Neutrinos in Physics and Astrophysics''(World Scientific, 1998); J. W. F. Valle, ``Gauge Theories and the Physics of Neutrino mass'', Prog. Part. Nucl. Phys. {\bf 26}, 91 (2001).
\bibitem{review} W.-M. Yao et al., J. of Phys. G {\bf 33}, 1 (2006).
\bibitem{grifols} J. A. Grifols, E. Masso, and S. Mohanty, Phys. Lett. B {\bf 587}, 184 (2004).
\bibitem{grimus} W. Grimus, M. Maltoni, T. Schwetz, M. A. Tortola, and J. W. F. Valle, Nucl. Phys. {\bf B 648}, 376 (2003).
\bibitem{joshipura} A. S. Joshipura and S. Mohanty, Phys. Rev. {\bf D 66}, 012003 (2002).
\bibitem{beacon} J. F. Beacom and P. Vogel, Phys. Rev. Lett. {\bf 83}, 5222 (1999).
\bibitem{pulido} J. Pulido and A. M. Mourao, Phys. Rev. {\bf D 57}, 7108 (1998).
\bibitem{montanino} D. Montanino, M. Picariello, and J. Pulido,  Phys. Rev. D {\bf 77}, 093011(2008).
\bibitem{kang} S. K. Kang, J. E. Kim, and J. S. Lee, Phys. Rev. {\bf D 60}, 033008 (1999).
\bibitem{Fukuda} Y. Fukuda et al., Phys, Rev. Lett. {\bf 81}, 1562 (1998).
\bibitem{lalakulich} O. Lalakulich, E. A. Paschos, and G. Piranishvili, Phys. Rev. {\bf D 74}, 014009 (2006).
\bibitem{Penner} G. Penner and U. Mosel, Phys. Rev. {\bf C 66}, 055212 (2002); M. Post, Ph.D. Thesis, Universitat Giessen, 2004.
\bibitem{ruso} T. Leitner, L. Alvarez-Ruso, and U. Mosel, Phys. Rev. {\bf C 74}, 065502 (2006).
\bibitem{nieves} E. Hernandez, J. Nieves, and M. Valverde, Phys. Rev. {\bf D 76}, 033005 (2007).

\end{references}
\end{document}